# Evolution of Glassy Carbon Microstructure: *In Situ* Transmission Electron Microscopy of the Pyrolysis Process

Swati Sharma,[1*] C.N. Shyam Kumar,[2,3] Jan G. Korvink,[1] Christian Kübel [2,4,5]

[1] Institute of Microstructure Technology, Karlsruhe Institute of Technology, Hermann-von-Helmholtz-Platz 1, 76334 Eggenstein-Leopoldshafen, Germany

[2] Institute of Nanotechnology, Karlsruhe Institute of Technology, Hermann-von-Helmholtz-Platz 1, 76334 Eggenstein-Leopoldshafen, Germany

[3] Department of Materials and Earth Sciences, Technische Universität Darmstadt, Alarich-Weiss-Straße 2, 64287 Darmstadt, Germany

[4] Helmholtz Institute Ulm, Helmholtzstraße 11, 89081 Ulm, Germany

[5] Karlsruhe Nano Micro Facility, Karlsruhe Institute of Technology, Hermann-von-Helmholtz-Platz 1, 76344 Eggenstein-Leopoldshafen, Germany

**Abstract**

*Glassy carbon is a graphene-rich form of elemental carbon obtained from pyrolysis of polymers, which is composed of three-dimensionally arranged, curved graphene fragments alongside fractions of disordered carbon and voids. Pyrolysis encompasses gradual heating of polymers at ≥ 900 $^o$C under inert atmosphere, followed by cooling to room temperature. Here we report on an experimental method to perform in situ high-resolution transmission electron microscopy (HR-TEM) for the direct visualization of microstructural evolution in a pyrolyzing polymer in the 500-1200 $^o$C temperature range. The results are compared with the existing microstructural models of glassy carbon. Reported experiments are performed at 80 kV acceleration voltage using MEMS-based heating chips as sample substrates to minimize any undesired beam-damage or sample preparation induced transformations. The outcome suggests that the geometry, expansion and atomic arrangement within the resulting graphene fragments constantly change, and that the intermediate structures provide important cues on the evolution of glassy carbon. A complete understanding of the pyrolysis process will allow for a general process tuning specific to the precursor polymer for obtaining glassy carbon with pre-defined properties.*

[*]**Corresponding author. Email: swati.sharma@kit.edu, Phone: +49 721 608 29317.**

The conversion of lithographically patterned polymers into glassy carbon (IUPAC name: glass-like carbon, also known as vitreous carbon) via pyrolysis is an effective method for transforming micrometer-scale structures into nano-scale owing to the characteristic dimensional shrinkage.[1-5] Consequently this process, which was conventionally used for the bulk production of glassy carbon, has gained considerable popularity among microsystem engineers.[6-12] A variety of miniaturized glassy carbon structures such as MEMS and NEMS components,[7-9] microelectrodes[10-12] and sensors[13, 14] fabricated by carbonization of patterned polymers have emerged as reliable and inexpensive alternatives to synthetic graphene based electronics. This increasing interest in glassy carbon has created a compelling need for a thorough understanding of its microstructure and properties pertinent to nano-scale structures. However, the widely accepted microstructural models of glassy carbon describing it to be composed of either (i) interconnected graphene ribbons with voids,[15] or (ii) cage-like graphene structures similar to fullerenes,[16] are primarily based on the characterization of commercially available glassy carbon, manufactured using industry-scale processes. While these models do substantiate the general physicochemical properties of glassy carbon, they are inadequate in explaining the microstructural variations arising from the fact that structures with a high surface-to-volume ratio feature different porosity and defect annealing patterns.[17]

Model (i) was first suggested by Jenkins and Kawamura in 1971,[15] while model (ii) has been reported in various studies by Harris (1997-2005),[16, 18-20] which is inspired by the work of Franklin (1951).[21] In Franklin's report[21] the term 'glassy carbon' was not used. She classified polymer-derived carbons as graphitizing and non-graphitizing and suggested that non-graphitizing carbons cannot be converted into crystalline graphite even at very high pyrolysis temperatures. Activated and glassy carbons fall into this category. In addition to these models, various aspects of microstructure and properties of bulk-manufactured non-graphitizing carbons are detailed in several books,[22-25] reviews,[26, 27] research articles[28-34] and respective cross-references. The driving force in most of these studies is the observation of (i) impermeability despite low density indicating the presence of closed pores, and (ii) crumpled graphene sheets with a predominantly turbostratic (misaligned basal planes) arrangement suggesting defects. In terms of conclusions, there is an obvious lack of consensus in these reports. One may also encounter occasional differences in the nomenclature or even in the general description of the physicochemical characteristics of the material. As a result, it is often difficult to directly compare the properties of a miniaturized glassy carbon structure fabricated, for example, to support a device application with the available information. One reason for this ambiguity is that glassy carbon is not a unique material. Its exact microstructure is known to be influenced by pyrolysis conditions,[12, 35, 36] chemical composition of the precursor polymer[37] and in the case of nano-scale structures, the forces applied during polymer-patterning.[8]

Pyrolysis encompasses thermochemical decomposition of polymers.[22] At initial heating stages (typically below 550 $^{o}$C), a carbonaceous backbone is formed,[26, 38] which serves as the skeleton for the ensuing glassy carbon. Pyrolytic carbon obtained in the 550-700 $^{o}$C temperature range is known to contain a high fraction of dangling bonds, which results a strong electron paramagnetic resonance (EPR) signal confirming the presence of active radicals.[35] Process temperatures > 700 $^{o}$C lead to C-C bond formation, development of short-range order, and an increase in the stacking thickness ($L_c$) followed by in crystallite diameter ($L_a$), as determined by various X-ray diffraction (XRD) and Raman spectroscopic studies.[39] Temperatures > 900 $^{o}$C induce further graphitization. Voids and in-plane non-six membered rings are partially annealed out as defects. However, some defects cannot be annealed even at very high temperatures, causing the material to remain non-graphitizing with its characteristic low density. The exact temperature of each pyrolytic transition, as well as the backbone structure are dependent on the chemical structure of the precursor polymer, and the collective rate of various parallel thermochemical reactions taking place within the pyrolyzing material.[36, 40, 41] Evidently, the formation and collapse of intermediate structures during pyrolysis can provide important cues on the resulting organization of the graphene fragments. Some such intermediates are short-lived or undergo major reconfiguration on cooling. As a result, it is essential to perform the microstructural analysis on the dynamic pyrolyzing material itself.

TEM is the most suitable tool for conducting such an investigation on a nano-scale sample. Other potential techniques such as high-temperature XRD or *in situ* Raman spectroscopy[41] utilize micrometer-scale samples, may suffer from peak-broadening due to constant bond-length fluctuations, and do not provide any visual data that can be directly correlated with an anticipated carbon nanostructure. However, one major drawback of TEM is the radiation damage, especially in the case of highly beam-sensitive materials such as carbon and polymers.[42-44] The threshold for atomic displacement (knock-on damage) caused by the direct collisions between the (beam) electrons and the nuclei of carbon atoms in graphene rings is just above 80 kV.[42] This implies that acceleration voltages > 80 kV may cause severe electron-beam induced damage to the sample,[42, 43] and thus introduce noticeable defects and/ or change the network configuration of the graphene sheets, rendering the material microstructurally transformed. Consequently, in order to obtain reliable TEM data, the imaging must be performed at low-voltages. Low-voltages are particularly important in the case of *in situ* analyses of a continuously changing material, where a clear distinction between thermochemically driven and beam-induced transformations is essential.

Here we report on an experimental approach for conducting low-voltage (80 kV) *in situ* TEM analysis of pyrolyzing SU-8 nanostructures directly patterned on to MEMS-based heating chips. Our hypothesis is that a complete understanding of glassy carbon microstructure entails an *in situ* investigation of the entire carbonization process. Particularly in the case of TEM analyses, acquiring sequential *in situ* data is the only way to determine if a projected structure is two- or three-dimensional. The results are compared with the two aforementioned models (graphene ribbons and fullerene-related microstructure). We also speculate that the microstructure of glassy carbon is strongly influenced by the size and morphology of the initial sample. Additionally, we propose that the microstructure of any given glassy carbon sample is a combination of the two previous models. For example, the graphene fragments feature random shapes and sizes rather than ribbons as suggested by Jenkins and Kawamura.[15] On the other hand, the material does exhibit a measurable graphitic stacking and inter-fragment links, instead of mainly discrete fragments trying to attain strongly folded or closed-cage morphologies, as it appears in the pictorial model reported by Harris.[16]

Two salient features of our experimental design, which can be extended to practically any precursor polymer, are the use of MEMS-based heating chips[45] and the low voltage TEM capabilities. Advances in low-voltage TEM (30-80 kV) have enabled the characterization of graphene-based materials without substantial beam damage while still maintaining a high resolution.[46, 47] Reported heating platforms have already been utilized for high-temperature *in situ* TEM imaging of single and few-layer graphene[45, 48, 49] and CNTs,[50, 51] understanding the nucleation mechanism of carbon nanostructures[52] as well as for *in situ* electrical biasing.[53, 54] However, to our knowledge the use of such chips has not been extended to deciphering the microstructure of glassy carbon. Importantly, direct sample patterning onto these chips bypasses the harsh TEM sample preparation steps, often involving extensive milling or focused ion beam (FIB) thinning that can cause significant stresses and ion implantation,[55] thereby altering the original microstructure.[56, 57] The use of silicon-based chips as substrates, probing SU-8 nanostructures, inert pyrolysis environment, and the availability to perform programmable pyrolysis up to 1200 °C with well-regulated heating and cooling rates, render this experimental scheme very close to the practiced carbon-MEMS/ NEMS fabrication process.[3, 4] Finally, we have taken utmost care of TEM data analysis. TEM images can be potentially misinterpreted owing to the lack of depth perception, *i.e.*, 2D projections of a 3D material.[58] Several nano-geometries that are likely to be confused with fullerenes or nanoparticles are discussed in detail. We also describe how *in situ* data on the evolution of such structures allows for a backward trace that alludes to their actual shapes.

**Results and Discussions**

*Sample morphology and imaging locations*

Different sample types that were probed by *in situ* TEM imaging are presented in **Figure 1**. Figure 1A is an optical micrograph of a SU-8 fiber patterned across three imaging windows of a heating chip. TEM micrographs of a freestanding cantilever-like structure, the edge of a fiber and a thin-film region are shown in Figure 1B, C and D, respectively.

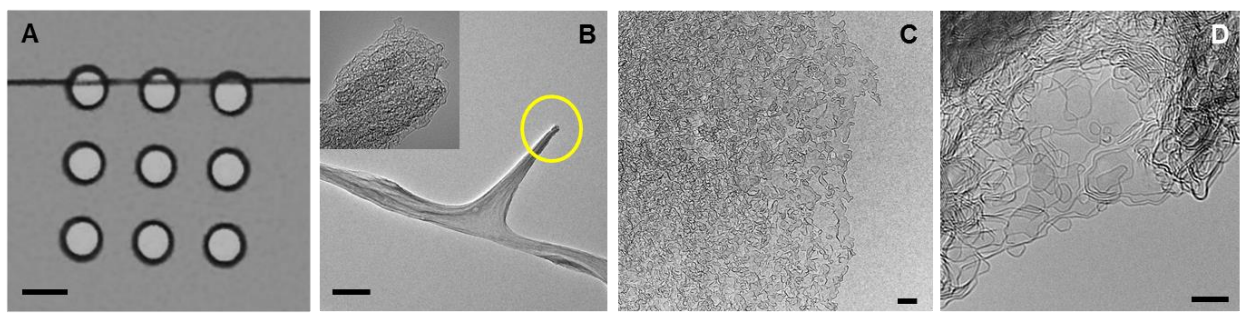

**Figure 1: Various SU-8 samples patterned on to heating chip.** (**A**) Optical micrograph of a SU-8 fiber patterned across three imaging windows. TEM images of (**B**) freestanding cantilever-like structure obtained by plasma etch of a fiber (Inset: highlighted tip region), (**C**) edge of a fiber, and (**D**) thin-film region at the edge of an imaging window. Scale bars: A: 10 μm, B: 200 *nm*, C: 10 *nm*, D: 5 *nm*.

Microstructural evolution of these structures during pyrolysis was analyzed to provide specific information on fullerene-like and other closed-cage geometries, migration and reorganization of the graphene fragments, graphitic stacking, fragment bifurcations and edge dynamics during pyrolysis. Structures with and without substrate were also compared for their post-pyrolysis shrinkage, which may be useful for microsystem engineers in determining the initial geometry of a pattern for nano-device fabrication. The substrate-supported fiber exhibited a unidirectional shrinkage (along *z*-axis), while the substrate-less structures (such as the one in Figure 1B) shrank isometrically.

*Post-pyrolysis glassy carbon*

TEM images shown in **Figure 2** were recorded at the edge of a SU-8 nanofiber pyrolyzed at 1200 $^o$C inside the TEM chamber. Three regions that resemble strongly folded or closed-cage structures (numbered 1-3) are shown in higher magnification to the left.

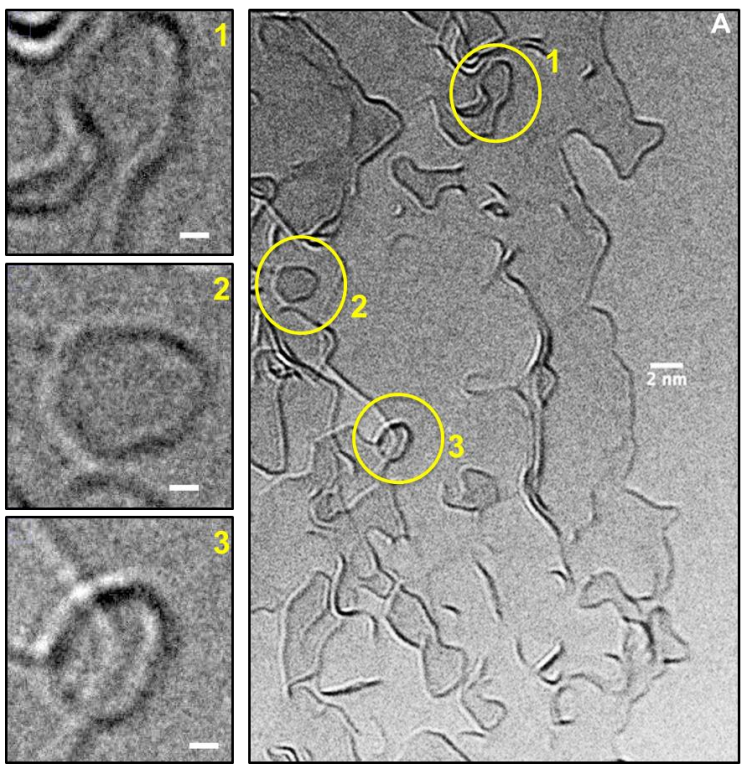

**Figure 2. Post-pyrolysis glassy carbon (pyrolysis temperature: 1200 $^o$C).** (A) TEM image taken at the edge of a substrate-supported glassy carbon fiber (scale bar: 2 *nm*). Highlighted regions (1-3) are shown in higher magnification (scale bars: 1 *nm*) to the left.

Region 1 represents a curved graphene sheet that seems to lead towards a closed-cage formation. However, this feature is most likely a projection of a graphene fragment spread in 3D, which may actually have a significantly different curvature than what appears in the 2D projection. The possibilities of its closing or unfolding remain open at higher annealing temperatures. The angular formation shown in Region 2 that contains near-120$^o$ bends, has a higher probability of representing a floating sheet than a completely closed, 3D structure. This feature is most likely a flat, disc-like graphene fragment. Region 3 gives the impression of a circular structure, but on a closer observation one can detect minor contrast variations on its periphery. This structure is probably a tilted graphene fragment, where the tilting is causing different defocus conditions at the edges. Similar structures are often mistaken for fullerenes or nanoparticles in the pre-manufactured glassy carbons, owing to the lack of their structural evolution data.

We elucidate this with simplified drawings in **Figure 3**, describing the various geometric possibilities for a projected pattern. Some of these schematics also facilitate discussions for the *in situ* TEM data detailed in the subsequent text. As can be observed in Figure 3A and B, the projected views representing a near-circular structure, and a set of concentric circles, can be rendered in various 3D shapes. Similarly, the illustration in Figure 3C could depict the bifurcation of a graphene fragment at a trigonal junction, or simply present a set of hexagons in different (*xy*) planes (corresponding TEM image in Figure 7). Figure 3D is the demonstration of a collective projection effect of various graphene fragments existing at different depths, which give the impression of intersections or crossovers, since only their edges are visible in a TEM image.

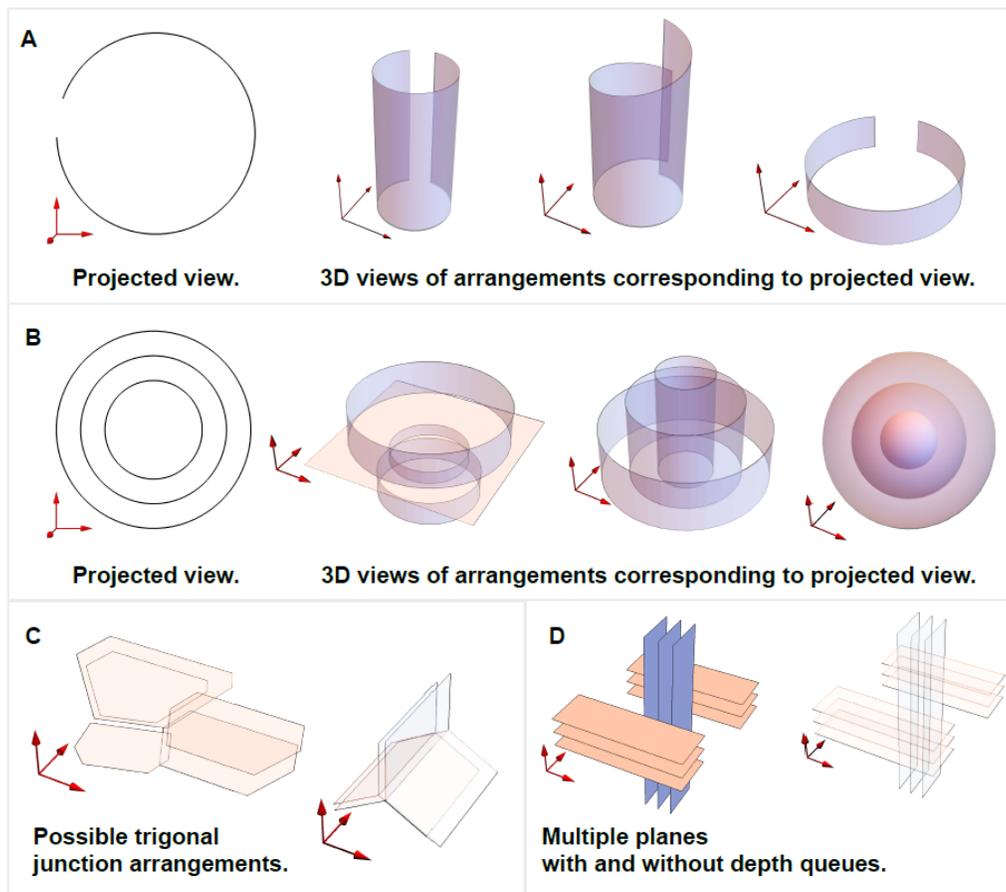

**Figure 3. Schematic representations of projected (2D) and 3D views of geometries typically observed in the TEM images of glassy carbon.** Projected-view and possible 3D views of apparently **(A)** near-circular geometry, and **(B)** concentric circles. **(C)** Possible graphene sheet arrangements at an apparent trigonal junction. **(D)** Multiple projected graphene edges present in different planes that appear to be intersecting. All schematics are for demonstration purposes and do not necessarily represent nanostructures present in glassy carbon.

*In situ observation of fullerene-like structures*

**Figure 4** is a compilation of detailed *in situ* TEM micrographs of a freestanding film during pyrolysis. The focus here is to evaluate the formation of fullerene-like structures with the help of the intermediates formed during pyrolysis. Prior to this, it is important to mention that the term 'fullerene-like' in this report is used specifically for carbon nanostructures that are compared with completely closed buckminsterfullerenes. In the literature, this term is occasionally used in a more general sense, *i.e.*, to describe a variety of graphene-based structures that contain pentagons/ heptagons or feature a high curvature. This has been cautiously avoided.

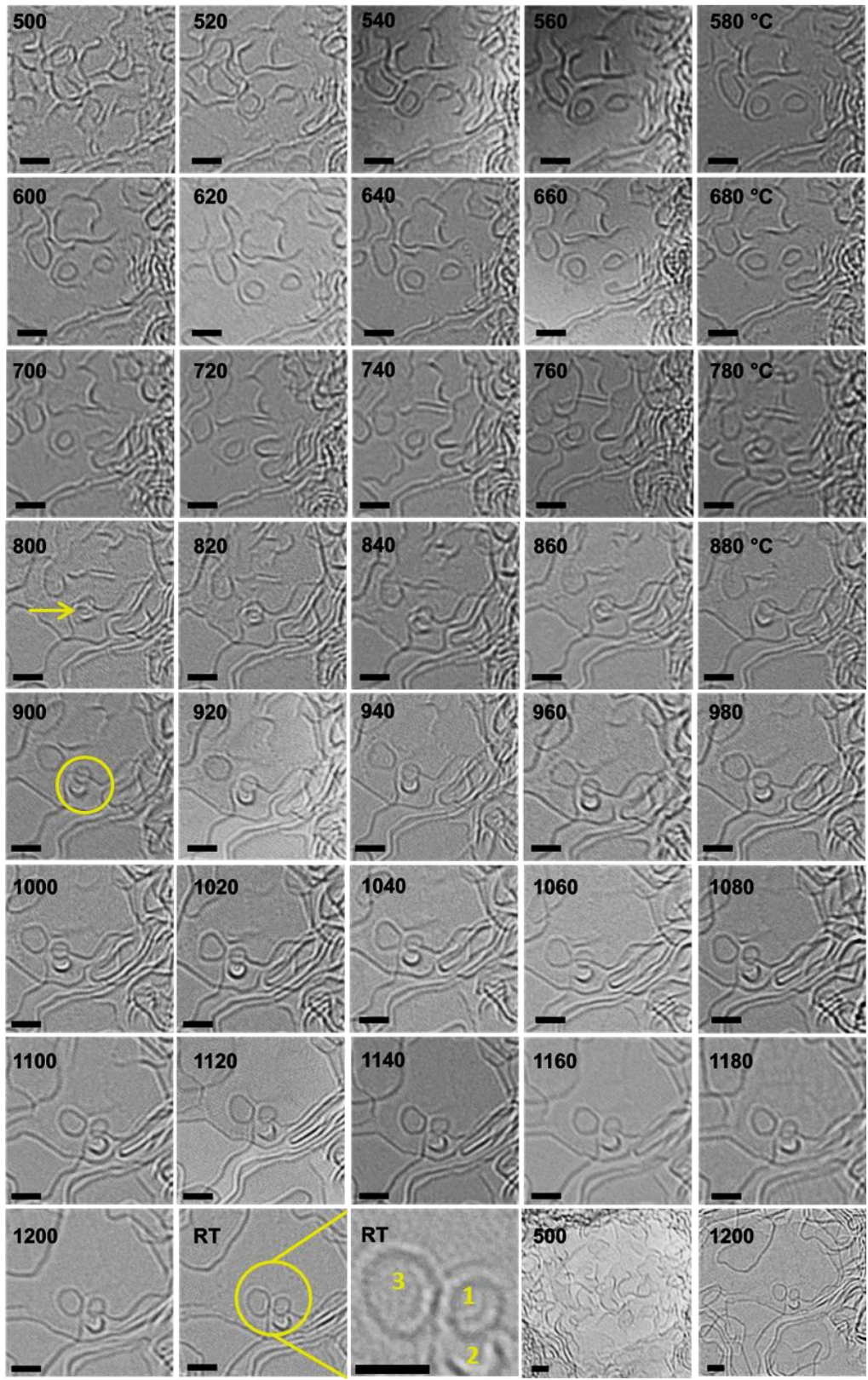

**Figure 4.** *In situ* **TEM images of a pyrolyzing SU-8 thin-film up to 1200 °C.** Pyrolysis temperature (in °C) at the time of image acquisition is displayed on each micrograph. All scale bars are 1 *nm*.

As can be observed in Figure 4, near-circular projections of carbon nanostructures are already traceable during the preliminary pyrolysis stages (see image at 520 °C). However, at these stages (typically up to 900 °C), the material undergoes a rapid reconfiguration. After pyrolysis (see RT images), 3 well-defined circular carbon nanostructures (labeled 1-3) are identified, which are further examined for being fullerenes using *in situ* structural evolution data.

Structures 1 and 2, which first appear around 800 °C (indicated by an arrow), display negligible change in shape or size up to 1200 °C. They slightly move apart, but do not make any attempt to attach to a neighboring fragment, despite a rapid rearrangement in the surrounding material. This structural stability supports the idea of a closed-cage formation. Importantly, the diameters of these two structures are approximately 0.7 *nm*, comparable with that of Buckminsterfullerenes. Even with this information, only structure-1 can be claimed as a fullerene. Minor shape and edge-contrast variations between 1080-1200 °C add uncertainty to defining structure-2. One cannot overrule the possibility that it is simply a protruding edge of another fragment. Structure 3 is slightly larger, which can either be a closed-cage (spherical) structure or a flat graphene flake resembling a disc. Without the knowledge of its evolution history, there is a high probability that it is misconstrued for a fullerene or a nanoparticle. However, on a careful backward trace one can witness the frequent distortion in its shape, as well as its migration, which corresponds to that of a floating disc-like fragment, not a spherical particle.

A similar structural development leading to the formation of closed-cage structures was observed at the tip of the cantilever-like structure shown in Figure 1B. Sequential *in situ* TEM images (100 °C intervals) recorded at this location are presented in **Figure 5** (A-H). Similar to the experiment shown in Figure 4, the formation of a stable closed-cage structure was only observed at around 900 °C (Figure 5D).

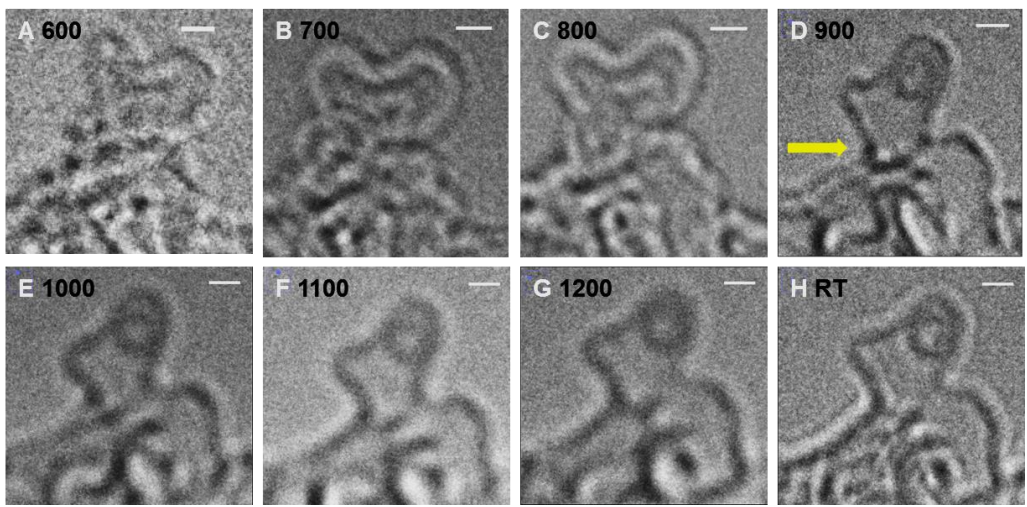

**Figure 5.** *In situ* **TEM micrographs recorded at the tip of a freestanding cantilever-like SU-8 structure.** The pyrolysis temperature (in $^{o}C$) at the time of image acquisition is indicated on each micrograph. All scale bars are 1 *nm*.

It is well-established that there is a sudden enhancement of glassy carbon's electrical conductivity and Young's modulus around 900 $^{o}C$.[4] From the micrographs in Figure 4 and 5, one obtains visual evidence of a prevailing microstructural change around 900 $^{o}C$. Similar to the micrographs shown in Figure 4, the closed-cage structure formed at the tip does not undergo any significant shape modification, despite constantly reorganizing surrounding material. These micrographs efficiently demonstrate how *in situ* imaging capabilities provide previously unknown essential information on the evolution of glassy carbon.

A comparison of *in situ* TEM data with the known physicochemical properties of low-temperature pyrolytic carbons (*e.g.* EPR characteristics[35]) suggests that the mobile graphene fragments with short-range order carry a large fraction of edge-radicals, and are constantly attempting to attain thermodynamically stable arrangements. A majority of fragments contain non-six membered rings, which cause them to curl and fold. As a result, the pyrolyzing material during the initial pyrolysis stages contains constantly migrating, curved and bent, floating graphene fragments with highly reactive edges. Occasionally, these mobile fragments form completely closed structures such as fullerenes. Some such defect-containing fragments also stack up, which is essentially turbostratic, since the defects cause misalignments in the basal planes and thus restrict a graphitic stacking (ABABA type).

Many structural defects are carried forward from the backbone structure[38] that are in turn influenced by polymer's chemical composition (distribution and nature of the C-heteroatom bonds) and the arrangement/entanglement of its chains. We speculate that some defects are also generated during pyrolysis. Generally, symmetric, six-membered ring based configurations are energetically favored in a graphene sheet. But in the case of a dynamic 3D material such as glassy carbon, it is possible that the surrounding matrix (composed of various reactive edges and already curved sheets) facilitates the formation of non-six-membered rings at a lower overall energy. Plausible reasons include steric hindrance, topological frustrations,[59, 60] or just a lack of neighboring atoms, *e.g.*, near the sample boundaries and surfaces.

*Separation and merger of graphene fragments*

**Figure 6** is a subcollection of the TEM micrographs shown in Figure 4 with an emphasis on the migration of small, floating graphene flakes during early pyrolysis stages.

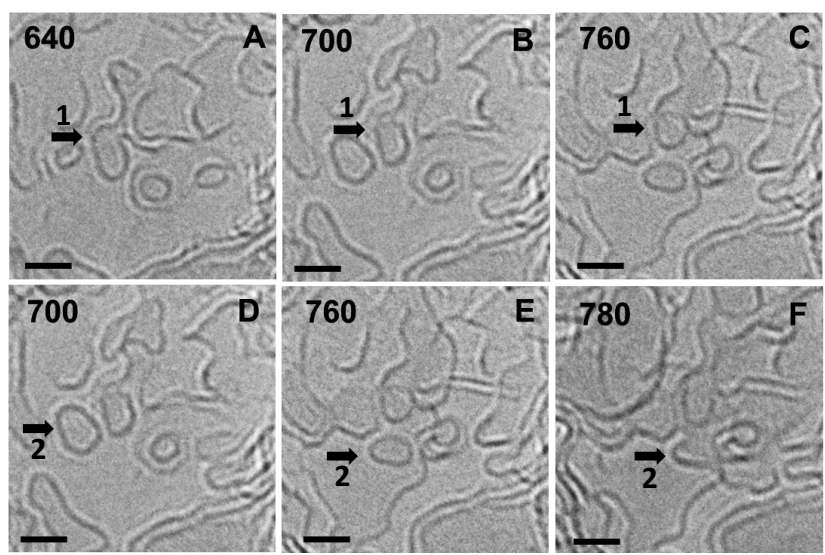

**Figure 6. Migration of small graphene flakes during pyrolysis. (A-C)** Separation of a circular flake from a larger graphene mass (arrow 1). **(D-F)** merger of a flake (which was absent from the field of view at 640 °C) into a neighbouring material at 780 °C (arrow 2). All scale bars: 2 *nm*.

In Figure 6A-C, the separation of a small fragment (diameter: < 2 *nm*) from a larger mass of graphene can be witnessed. Initially, a neck-like structure is formed at an unstable region around 640 °C, which subsequently develops into an independent flake at higher pyrolysis temperatures. The micrographs in Figure 6 D-F represent the merger of another flake, which entered the field of view of these micrographs at around 700 °C, into a larger fragment. Notably, the two transitions take place in the same temperature range and are perhaps interdependent. From these micrographs it is clear that the mobile graphene fragments are capable of completely merging or separating out, which would cause a sudden change in their size along the basal plane ($L_a$). Such migration patterns are more prominent during the early pyrolysis stages, likely due to the instabilities caused by their higher mobility.

*Multi-layer graphene and trigonal junctions*

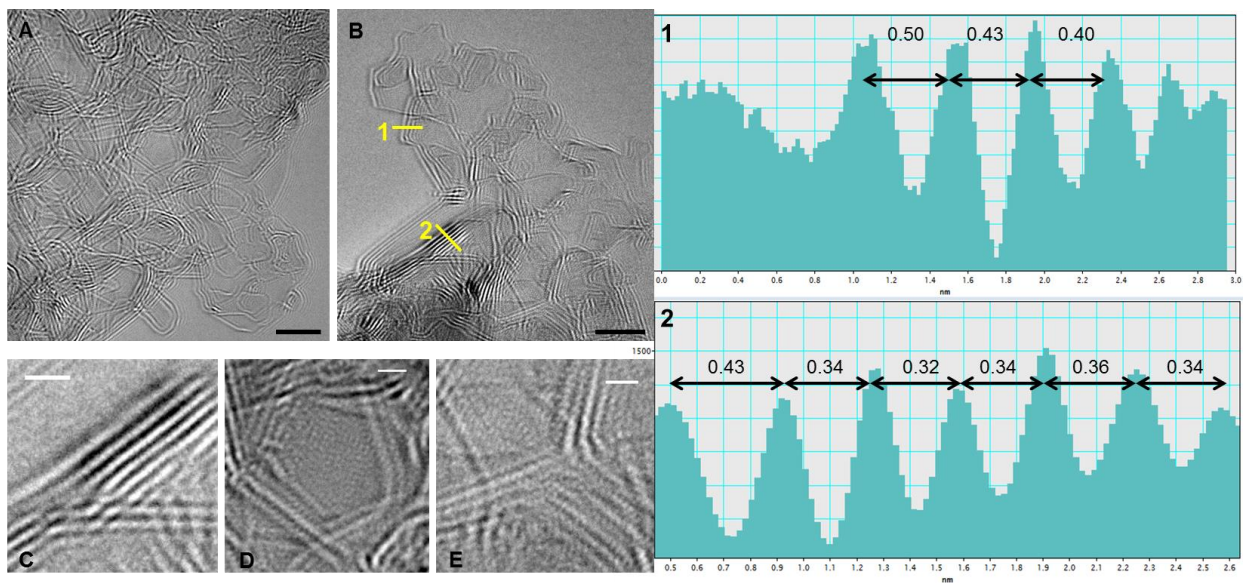

**Figure 7. TEM micrographs and inter-layer separation in a relatively thick (3D) glassy carbon sample.** **(A, B)** TEM images of post-pyrolysis glassy carbon featuring multiple graphene layers. **(C-E)** Selected regions from A and B that give the impression of **(C)** intersections or crossovers, **(D)** randomly shaped voids, and **(E)** a trigonal junction. **(1, 2)** Intensity profiles of regions marked 1 and 2 in (B). Scale bars: A, B: 5 *nm*; C-E: 1 *nm*.

Figure 7A and B are the micrographs recorded for a relatively thick fiber sample (multiple graphene fragments in all directions) after pyrolysis at 1200 °C. Similar micrographs have been reported by various researchers, since common glassy carbon samples are much thicker than the few-layer graphene. Here various graphene fragments present at different depths are projected on the image plane such that they seem to intersect or cross. One such location is magnified in Figure 7C. Such apparent intersections, previously denoted as a glassy carbon motif,[26] are merely a visual effect. Various such geometries can be traced in the parent images (Figure 7A, B). Since only the edges of a fragment are visible in TEM, the planes also appear to be more densely packed compared to their actual configuration (see Figure 3D). Figure 7D is yet another projection-derived geometry. Similar features have been confused for voids or nanoparticles in the literature due to their seemingly closed shape and ~5 *nm* size.

In order to confirm whether the dense edges represent $L_c$ or are merely visual effects caused by multiple projected planes, one can acquire their intensity profiles and determine the inter-layer distance. Region 1 in Figure 7A most likely contains few-layer graphene. We obtained its intensity profile (shown to the right), which yielded a distance between 0.4 and 0.5 *nm*. This is significantly higher than the characteristic graphite (0.335 *nm*) or turbostratic graphene (0.336-0.344 *nm*[61]) arrangements. The intensity profiles of another location (marked Region 2) featured an inter-layer separation indicative of a nearly graphitic stack within the measurement error.

We also observed that the graphene fragments often stack up on top of each other with a certain offset, rather than being perfectly edge-matched (also see Figure 8J). Due to the edge-only projections, the top view created by such an offset can be confused for $L_c$. Intensity profiles can provide some clarity, but only in the case of few-layer graphene. The identification of the orientation of the stacks becomes increasingly difficult with an increase in the sample thickness. Nonetheless, from these intensity profiles we can deduce that there is definitely some graphitic stacking and inter-fragment bonds in glassy carbon, which is often manifested by an increase in $L_c$ in the XRD patterns. Discrete fragments with no or very little stacking, as proposed by Harris, would also not justify glassy carbon's good electrical conductivity.

In Figure 7E a trigonal junction can be observed, which could be an example of the bifurcation of graphene sheets with 120° angles, featuring a confluence-like geometry. It is also possible that it is an illusion created by various hexagonal graphene fragments lying in different planes. Both possibilities are schematically represented in Figure 3C. It has been proposed that such junctions may contain tetrahedral geometries or diamond-like carbon (DLC).[26, 33] Since the nano-scale tetrahedral carbons are only stable up to 700 °C, such claims have also been repeatedly rejected.[16, 26] Extreme sample preparation conditions, or beam-damage can potentially introduces some $sp^3$-hybridized atoms, especially on the surface of glassy carbon. If the analysis is based solely on TEM, it is also possible that the trigonal geometries (such as in Figure 7E) rotated from the image plane are projected at an angle other than 120° and are therefore confused for DLC. In any case, the presence of DLC cannot be evaluated by the reported TEM experiment due to the complex projection effects in the micrographs.

The micrographs in Figure 7 are visually different from those recorded for the thin-films (Figures 4-6). This is a clear evidence of the influence of sample size and geometry on the resulting microstructure. A relatively thick sample contains a large fraction of angular bends (several 120$^o$ angles identified in Figure 7A, B), which is not the case for the thin films. Multilayered structures allow for 3D interactions between various developing fragments. As a result, the fragments can spread and bond across different planes, and the stacked crystallites can orient themselves in any direction. On the other hand, graphene fragments in thin-films tend to expand primarily in the *xy* plane, since any out-of-plane protrusions would require a very high surface energy.

It is evident from the data shown in Figures 4-7 that the material (above 900 $^o$C) is composed of randomly shaped graphene fragments of highly variable sizes, interconnected in a complex manner, rather than long and narrow, frequently bifurcating ribbons. These fragments do contain non-six-membered carbon species that introduce curvature in the fragments resulting in occasional formation of nanoparticles or fullerenes that are typically 0.7-2.0 *nm* in diameter. Slightly larger nanostructures are most likely floating flakes or projection effects. The term ribbon gives the impression that graphene sheets in glassy carbon feature a much greater length compared to their stack thickness, and also that the width of these so-called ribbons is more or less uniform along their entire length. Such a proposition is based on the hypothesis that the polymeric backbone serves as the nucleation point or the central axis for the growth of a graphene fragment.[26] Since the backbone is derived from long polymer chains, a graphene sheet originating from it would look like a ribbon, fibril or chain. The expansion of such ribbons was suggested to be restricted by the edge atoms featuring valance angle variations (non-$sp^2$ hybridizations), or by steric hindrance due to the neighboring ribbons.[22] Based on the *in situ* TEM data, we conclude that such ribbon-like graphene growth is a highly unlikely scenario, since the developing fragments contain reactive edges in all directions with no restriction to spread. Additionally, smaller flakes are quite mobile and can attach to other fragments. The separation and merger of graphene fragments indicates that instead of always expanding, they are constantly rearranging to achieve a more stable overall structure. The assumption that several stacked ribbons exhibit almost identical geometries is also improbable. In the reported work it was difficult to find even two identical fragments. The ribbon model is often supported by the XRD data that indicates an increase in $L_c$ and $L_a$, before and after the pyrolysis temperature of ~1000 °C. Here one needs to be careful, since $L_a$ is defined as the expansion of the graphene basal plane that can occur in any, or simultaneously in all directions along the edges (*e.g.* radial growth). Often the ribbon or fibril models assume the increase in $L_a$ to be unidirectional.

In the model proposed by Harris, the change in $L_c$ has been completely overlooked and the material is believed to contain a very high fraction of strongly folded or closed-cage structures. Most of these structures, including the relatively large ones (5-10 *nm*), are considered fullerene-like. Such structures may indeed be complex 2D projections from multiple planes, 2D (disc-like or hexagonal) flakes, or the top-view of graphene layers arranged with an offset. A clear explanation of what is considered fullerene, and what is 'fullerene-like', is also necessary for a better understanding of this model.

*Graphene edge dynamics during glassy carbon evolution*

Finally, in **Figure 8** the patterns that evolve at the graphene edges during pyrolysis are revealed. One can witness a continuous transformation of these edges, present at an unhinged boundary of a thin-film sample, from a highly disordered structure to well-defined facets similar to those observed in synthetic graphene nanoribbons.[62] The 3-layer (or 3-stack) arrangement at this dynamic protruding edge is clearly detectable around 1000 °C. Evidently, the edges are highly dynamic as they undergo continuous reorganization. Their curvature changes constantly and the structure ultimately acquires a symmetry pattern (which could be zig-zag or armchair at a molecular level) with bends close to 120°. Interestingly, the edges become angular, slightly rounded, and then angular again between 1000 °C to RT, confirming that during pyrolysis ramp-up, the material continues to reorder even at higher temperatures. A cool-down at any given temperature leads to an energetically favorable arrangement. Also noticeable is the frequent change in the interlayer separation, which clearly indicates that these micrographs feature the top view of graphene layers rather than the side-view or $L_c$ (see schematic in Figure 8J).

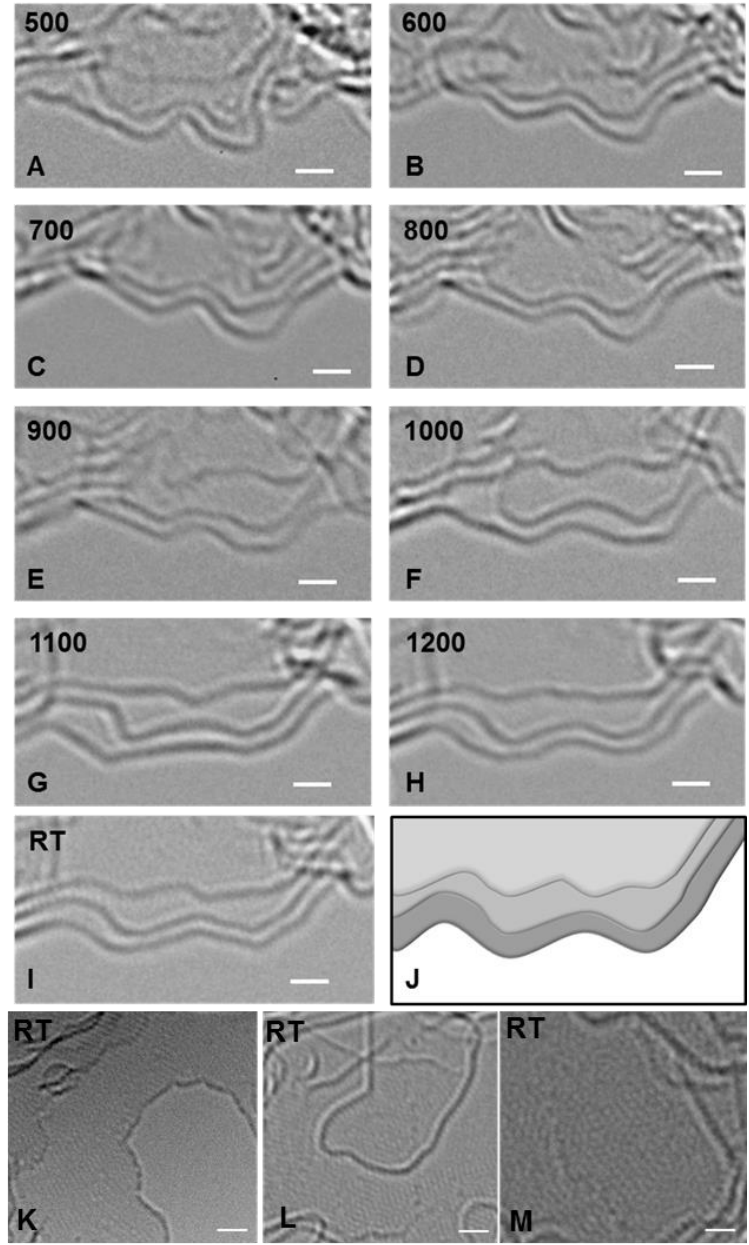

**Figure 8. Graphene edge dynamics during glassy carbon evolution. (A-J)** A developing 3-layer (or 3-stack) graphene structure. (**L,M,N**) Defect-containing graphene edges and grain boundaries in glassy carbon. All scale bars are 1 *nm*.

Glassy carbon contains a considerable fraction of reactive, defect-containing edges that are responsible for the high degree of fragment mobility. In Figure 8 K-M, the slightly out of focus edges can be observed, which appear coarse and discontinuous. These defective edges may contain open rings, dangling bonds, voids, or other intermediate structures, in addition to non-six membered rings. Here, the reactive edge atoms have the option of forming σ-bonds with a neighboring flake, or of attaching with the upper (or lower) layer by π-bonds. An in-plane bonding of the two defective edges could also lead to grain boundaries[63] that can be partially annealed out at higher temperatures.[64]

**Conclusions**

By comparing the data from all reported experiments we conclude that fullerenes, strongly curved graphene sheets, and small 2D graphene flakes, co-exist with relatively much larger, stacked (< 10 layers) and interconnected graphene fragments of highly variable sizes and shapes in glassy carbon. The fraction and distribution of fullerenes and fullerene-like structures, as well as the extent and nature of 3D inter-fragment bonding are strongly dependent on the surface area of the sample, in addition to the chemical structure of the precursor polymer. Microporosity in glassy carbon should not be completely attributed to fullerene-like structures. Arbitrarily shaped voids caused by inter-fragment bonding across multiple planes also contribute to material's low density. These random voids are more prevalent in 3D samples as opposed to the few-layer graphene version of glassy carbon. The fragment interconnects contain both σ and π bonds. Some of them may be highly strained resulting in a range of C-C bond-lengths in glassy carbon. Inherent non-six-membered rings, which are also responsible for a predominantly turbstratic arrangement, further contribute to bond-length variations. Graphene fragments in pyrolytic carbon are not always expanding. Occasionally smaller flakes separate from, or merge into the larger ones, due to local instabilities, especially at low pyrolysis temperatures.

The reported *in situ* TEM imaging protocol can be extended to practically any polymer in order to determine the exact microstructure of a specially designed glassy carbon. A complete understanding of the pyrolysis also allows for process tuning, for example, the change in dwell time or temperature ramp-up specific to the precursor polymer for obtaining the material with pre-defined properties. Various nanostructures in glassy carbon, for example, disc-like flakes as small as 2 *nm*, can only be deconvoluted with their complete evolution history, which is only possible with low-voltage TEM at the nano-scale. Other commonly used characterization techniques only yield the average values of material's properties and thus limit the microstructural information. The integration of low-voltage HR-TEM with the scanning-TEM can be potentially employed for obtaining more detailed information regarding glassy carbon microstructure in the future.

Glassy carbon is not only an excellent engineering material, it is also a platform suitable for understanding the origin of graphene from a hydrocarbon, and the behavior of graphene fragments in a dynamic 3D environment. There are numerous possibilities for theoretical studies that would provide further information on the minimum energy configurations, despite strained bonds. Mathematical simulations can also shed light on the exact mechanism of fullerene formation in glassy carbon. Our current experiments were limited to 1200 °C due to the chip design, which can be further improved for obtaining microstructural data at higher temperatures.

**Methods**

TEM experiments were carried out on an aberration corrected FEI Titan 80-300 microscope operated at 80 kV, equipped with an Aduro 200 (Protochips Inc.) sample holder. SU-8 (a photopatternable thermosetting phenol-formaldehyde resin commonly used for carbon-MEMS fabrication;[65] procured from MicroChem) fibers were manually placed onto Aduro E chips (Protochips Inc.) such that at least one fiber could be observed through the imaging window. Fibers were UV cross-linked and where necessary, further thinned by oxygen plasma etching prior to pyrolysis. Longer etch times resulted in fiber breakage at the center, thus yielding a cantilever-like structure (freestanding beam anchored at one end). Thin films were obtained when the patterned fiber barely touched the imaging window, allowing for extremely small quantities of SU-8 to be imaged. Chips were then placed in the Aduro 200 TEM holder and were heated inside the TEM chamber (pressure: ~$10^{-7}$ torr) at 5 °C/minute ramp rate up to 1200 °C, followed by cooling at a rate of 10 °C/ minute. Images were recorded at 20 °C intervals in the 500-1200 °C range. The beam was only turned on during imaging.

Schematics of the top and cross-sectional views of a heating chip and a digital photograph of the accompanying TEM holder are provided in **Figure 9**. The chip consists of a freestanding ceramic membrane coated with a thin amorphous silicon nitride (SiN) film with patterned holes as imaging windows. The ceramic membrane serves as the heating element and the SiN membrane (thickness: 50 *nm*) as the sample support.[66] Chips with and without the SiN membrane were used for substrate-attached and freestanding samples respectively.

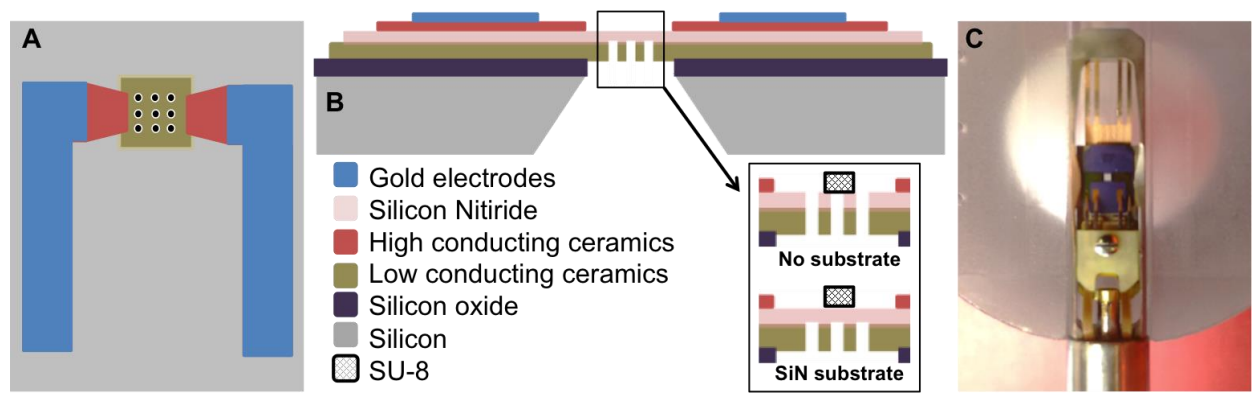

**Figure 9. Protochips devices used for *in situ* heating.** **(A)** Top and **(B)** cross-sectional view of a heating chip showing imaging windows (diameter: 9 μm) connected to gold contact pads.[66] **(C)** Aduro 200 *in situ* heating holder with mounted chip.

**Data availability:** Detailed TEM images (as recorded) are available from SS on request.

**Author contributions:** All authors contributed to the conceptual design and manuscript preparation. SS prepared the samples, interpreted the TEM data and correlated it to the glassy carbon microstructure. CNSK and CK facilitated and conducted low-voltage TEM and contributed to data interpretation.

**Acknowledgements:** SS sincerely thanks Prof. SS Bukalov, Russian Academy of Science, Moscow, for useful discussions on glassy carbon microstructure. SS and JGK acknowledge financial support from the Ministry of Science, Research and Arts, BW, Germany via Az:33-7533-30-20/3/3, HEiKA Center FunTECH-3D. CNSK is thankful to the German Academic Exchange Service (DAAD) for a Ph.D. scholarship.

**Competing interests:** Authors declare no competing interests.